\documentclass[preprint]{emulateapj}
\usepackage{epsf,psfig}
\begin{document}

\title{The Impact of Galaxy Formation on the Sunyaev-Zeldovich Effect of Galaxy Clusters}

\author{Daisuke Nagai}

\affil{Department of Astronomy and Astrophysics, \\ 
Kavli Institute for Cosmological Physics,\\ 5640 S. Ellis Ave., The
University of Chicago, Chicago IL 60637\\ e-mail: {\tt
daisuke@oddjob.uchicago.edu}}

\begin{abstract}  

We study the effects of galaxy formation on the Sunyaev-Zel'dovich
effect (SZE) observable-mass relations using high-resolution
cosmological simulations. The simulations of eleven individual
clusters spanning a decade in mass are performed with the
shock-capturing Eulerian adaptive mesh refinement N-body+gasdynamics
ART code.  To assess the impact of galaxy formation, we compare two
sets of simulations performed in an adiabatic regime (without galaxy
formation) and those with several physical processes critical to
various aspects of galaxy formation: radiative cooling, star
formation, stellar feedback and metal enrichment.  We show that a SZE
signal integrated to a sufficiently large fraction of cluster volume
correlates strongly with its enclosed mass, independent of details of
gas physics and dynamical state of the cluster.  The slope and
redshift evolution of the SZE flux-mass relation are also insensitive
to processes of galaxy formation and are well characterized by a
simple self-similar cluster model.  Its normalization, on the other
hand, is significantly affected by gas cooling and associated star
formation.  Our simulations show that inclusion of these processes
suppresses the normalization by $\approx 30-40\%$.  The effect is due
to a decrease in gas mass fraction, which is offset slightly by an
increase in gas mass-weighted temperature. Gas cooling and star
formation also cause an increase in total mass and modify the
normalization by a few percent.  Finally, we compare the results of
our simulations to recent observations of the SZE scaling relations
obtained using 36 OVRO/BIMA SZE$+${\it Chandra} X-ray observations.
The comparison highlights the importance of galaxy formation in
theoretical modelling of clusters and shows that current generation of
simulations produce clusters with gross properties quite similar to
their observed counterparts.

\end{abstract}

% User-supplied List of keywords.


%\altaffiltext{1}{Department of Astronomy and
%Astrophysics, Kavli Institute for Cosmological Physics, 5640 South
%Ellis Ave., The University of Chicago, Chicago, IL 60637}
%\altaffiltext{2}{{\tt daisuke@oddjob.uchicago.edu}}

%----------------------
\section{Introduction}
\label{sec:intro}
%----------------------

The Sunyaev-Zel'dovich effect (SZE) is a potentially powerful
observational tool for cosmology.  It is a small distortion in the
cosmic microwave background (CMB) spectrum caused by scattering of CMB
photons off a distribution of high energy electrons in dense
structures such as clusters of galaxies
\citep{sunyaev_etal70,sunyaev_etal72}.  This effect has the unique
property that its signal is independent of redshift, making it
particularly well suited for deep cluster surveys
\citep[e.g.,][]{holder_etal00,weller_etal02}. The next generation of
SZE instruments, such as the South Pole Telescope (SPT) and the
Atacama Cosmology Telescope (ACT), should be capable of mapping a
fairly large portion of the sky and finding a large number ($\gtrsim
10^4$) of clusters out to high-redshift.  Such large and homogeneous
sample of galaxy clusters will enable direct and precise measurements
of their number density as a function of redshift, and the expected
survey yields will be sufficient to provide one of the most powerful
constraints on the nature of dark energy
\citep{wang_etal98,haiman_etal01}.

To realize the full statistical power of the upcoming SZE surveys,
however, systematic uncertainties would have to be controlled at a
level comparable to statistical uncertainties.  One of the main
sources of the systematic uncertainties lie in the relation between
the SZE observable and cluster mass as a function of redshift.  Making
this connection is important because the cluster mass is not directly
observable.  For the future SZE surveys, the requirement is to control
systematic uncertainties in the SZE observable-mass relations to
better than $\sim$5\% at all redshift
\citep[e.g.,][]{carlstrom_etal02}.

This poses a serious challenge to both observers and theorists. To
date, observational studies of the SZE scaling relations have been
performed using two largest datasets of SZE measurements obtained by
the OVRO/BIMA cm-wave imaging experiment
\citep{cooray_etal99,mccarthy_etal03,laroque05} and multiband SuZIE
experiment \citep{benson_etal04} along with X-ray observations.
Analyzing a sample of 36 clusters observed with both the OVRO/BIMA SZE
imaging and {\it Chandra} X-ray observations, \citet{laroque05} showed
that there are tight correlations between the observed SZE flux and
X-ray temperature and cluster masses.  The observed regularity of the
SZE effect of clusters is encouraging news, but further progress is
clearly needed for the future SZE surveys.  The observational
situation is expected to improve rapidly with the advent of a number
of dedicated SZE survey instruments, which will dramatically increase
the sample size and the number of low-mass clusters.

On a theoretical side, a number of groups have studied the SZE scaling
relations using semi-analytic models
\citep{verde_etal02,mccarthy_etal03} and cosmological simulations
\citep{metzler_etal98,white_etal02,daSilva_etal04,diaferio_etal05,motl_etal05}.
Motivated by results of X-ray observations in a past decade
\citep[see][for a review and references therein]{voit_etal05}, recent
studies have focused on studying the effects of non-gravitational
physical processes, including gas cooling, star formation and energy
feedback, on the SZE scaling relations.  One of the main results is
that a SZE signal integrated to a sufficiently large fraction of
cluster volume is an extremely good proxy for its enclosed mass,
independent of details of gas physics and dynamical state of a cluster
\citep[see e.g.,][]{motl_etal05}.  The slope and the redshift
evolution of the SZE scaling relation also appear to be insensitive to
details of cluster gas physics \citep[e.g.,][]{daSilva_etal04}.  While
these results are encouraging for cosmological applications, these
previous studies have focused on simulating a large number of
clusters, and the resolution was inevitably limited to capture
relevant cluster physics. As such, the impact of galaxy formation on
normalization of the SZE scaling relations has not yet converged among
different simulations.  It is therefore important to push theoretical
modelling of SZE scaling relations and check previous results using
higher-resolution cluster simulations.

In this paper, we present such study using high-resolution
cosmological simulations of cluster formation.  Although the statistic
is limited, our cluster sample spans over a decade in mass and provide
a good leverage on scaling relations.  The mass resolution of our
simulations is more than an order of magnitude higher than that in
previous studies.  This work is therefore complimentary to the
previous studies in a literature.  Using these simulations, we study
the impact of gas cooling and star formation on the SZE scaling
relations, including their normalization, slope and redshift
evolution.  To test the results of our simulations, we compare our
results to recent observations of the SZE scaling relations based on a
sample of 36 clusters obtained using the OVRO+BIMA SZE and {\it
Chandra} X--ray telescopes \citep{laroque05}.

The paper is organized as follows. In \S~\ref{sec:model} we define
observational quantities and present relevant scaling laws predicted
by a self-similar cluster model.  We describe simulations presented in
this paper in \S~\ref{sec:sim} and present results and comparisons to
previous studies and recent observational results in
\S~\ref{sec:results}.  Finally, in \S~\ref{sec:discussion} we discuss
our conclusions and their implications for SZE cluster surveys.

%------------------------------
\section{Theoretical Framework}
\label{sec:model}
%-----------------------------

\subsection{Thermal Sunyaev-Zel'dovich Effect}

The thermal SZE is a distortion in the CMB spectrum produced by the
inverse Compton scattering of CMB photons off free electrons in dense
structures such as clusters of galaxies. For a given line of sight, a
change in the CMB specific intensity caused by the thermal SZE at a
frequency $\nu$ is given by $\Delta I_{\nu}/I_{\rm CMB}=f_{\nu}(x)
g_{\nu}(x) y$.  The dimensionless comptonization parameter $y$ is
defined as,
\begin{equation}
y \equiv \frac{k_B \sigma_T}{m_e c^2} \int n_e(l) T_e(l) dl.
\end{equation}
where $n_e$ and $T_e$ are the number density and temperature of
electrons, $m_e$ is the electron rest mass, $c$ is the speed of light,
and $\sigma_T$ is the Thomson cross-section.  There are several
frequency dependent factors, including $I_{\rm CMB}=2h\nu^3/c^2
(e^x-1)^{-1}$, $f_{\nu}(x)=[x(e^x+1)/(e^x-1)-4](1+\delta_{\rm
SZE}(x,T_e))$ and $g_{\nu}(x) = x^4 e^x/(e^x-1)^2$, where $\delta_{\rm
SZE}(x,T_e)$ is the frequency dependent relativistic correction and $x
\equiv h\nu/k_BT_{\rm CMB}$.  The corresponding change in the CMB
temperature is given by $\Delta T_{\nu}/T_{\rm CMB}=f_{\nu}(x) y$.  In
the Rayleigh-Jeans limit ($\nu \ll 200$GHz), $\Delta T_{\nu}/T_{\rm
CMB}= -2y$ and $\Delta I_{\nu}=(2k_B \nu^2/c^2) \Delta T_{\nu}$.

Let us now consider the SZE signal arising from a cluster located at
redshift $z$.  The SZE flux integrated over a solid angle of
observation $d\Omega$ is given by $\Delta S_{\nu} = \int \Delta
I_{\nu} d\Omega = I_{\rm CMB} f_{\nu}(x) g_{\nu}(x) Y$, where $Y$ is
the integrated Compton-y parameter defined as
\begin{equation}
\label{eq:Y}
Y \equiv \int_{\Omega} y d\Omega = \frac{1}{d_A^2(z)} \left(\frac{k_B
\sigma_T}{m_e c^2}\right) \int_V n_e T_e dV,
\end{equation}
and $d\Omega = dA/d_A^2(z)$ is the solid angle of the cluster
subtended on the sky, $d_A(z)$ is the angular diameter distance to the
cluster, $dA$ is the area of the cluster on the sky, and $dV$ is the
cluster volume. Because $Y$ depends on a distance, we will work with
the intrinsic thermal SZE signal, defined as
\begin{equation}
\label{eq:Yint}
Y^{\rm int} \equiv Y d_A^2(z) \propto f_{\rm gas} M T_m.
\end{equation}
Note that the integrated SZE flux is linearly sensitive to gas mass
$M_{\rm gas}=f_{gas}M$ and mass-weighted temperature $T_m$, where
$f_{gas}$ is the gas mass fraction and $M$ is the total cluster mass.

\subsection{Self-Similar scaling relations}

In the absence of cooling and heating processes, clusters are expected
to scale self-similarly \citep{kaiser_etal86}.  The self-similar model
predicts that the temperature of the gas scales with the cluster mass
as
\begin{equation}
\label{eq:mt} 
M \propto T^{3/2} E^{-1}(z)
%T \propto M_{h}^{2/3} [\Delta_{c}(z) E^2(z)]^{1/3}
\end{equation} 
where $M \equiv 4\pi r^3_{\Delta} \Delta_{c} \rho_{crit}/3$ is a halo
mass enclosed within $r_{\Delta}$, defined as a radius of spherical
volume within which the mean density is $\Delta_{c}$ times the {\it
critical density}, $\rho_{\rm crit}$, at that redshift
\citep{bryan_etal98}.  $E(z)$ is the redshift-dependent Hubble
parameter, defined as $H(z)=100hE(z)$ km s$^{-1} {\rm Mpc}^{-1}$, and
it is given by $E^2(z)=\Omega_M(1+z)^3+\Omega_{\Lambda}$ for a flat
cosmology.

Inserting Eq.~\ref{eq:mt} into Eq.~\ref{eq:Yint}, we obtain SZE
scaling relations predicted by the self-similar model,
\begin{equation}
\label{eq:Yscale}
Y^{\rm int} \propto \left \{ \begin{array}{ll}
	        f_{\rm gas} M^{5/3} E^{2/3}(z) \\
	        f_{\rm gas} T^{5/2} E^{-1}(z) \\
                      \end{array}
             \right.
\end{equation}

%----------------------
\section{Simulations}
\label{sec:sim}
%----------------------

In this study, we analyze high-resolution cosmological simulations of
eleven cluster-size systems in the ``concordance'' flat {$\Lambda$}CDM
model with $\Omega_{\rm m}=1-\Omega_{\Lambda}=0.3$, $\Omega_{\rm
b}=0.04286$, $h=0.7$ and $\sigma_8=0.9$, where the Hubble constant is
defined as $100h{\ \rm km\ s^{-1}\ Mpc^{-1}}$, and $\sigma_8$ is the
power spectrum normalization on $8h^{-1}$~Mpc scale.  The simulations
were done with the Adaptive Refinement Tree (ART)
$N$-body$+$gasdynamics code \citep{kravtsov99, kravtsov_etal02}, an
Eulerian code that uses adaptive refinement in space and time, and
(non-adaptive) refinement in mass \citep{klypin_etal01} to reach the
high dynamic range required to resolve galaxy-size halos formed in
self-consistent cosmological cluster simulations.

To set up initial conditions we first ran a low resolution simulation
of $80h^{-1}$~Mpc and $120h^{-1}$~Mpc boxes and selected eleven
clusters with mass ranging from $M_{500c}\approx 3.5\times10^{13}$ to
$9\times 10^{14}h^{-1}{\ \rm M_{\odot}}$.  Table~\ref{tab:sim} lists
properties of clusters at the present epoch. The perturbation modes in
the Lagrangian region corresponding to a sphere of several virial
radii around each cluster at $z$=0 was then re-sampled at the initial
redshift of $z_i=25$ for three most massive clusters (CL1-3) and
$z_i=49$ for remaining eight clusters in the sample.  For the three
most massive clusters we have resampled radius of $3R_{\rm
vir}$($z$=0), while for the rest of the clusters the resampling sphere
had radius of $5R_{\rm vir}$($z$=0), where $R_{\rm vir}$($z$=0) is the
virial radius enclosing overdensities of $\Delta_{\rm vir}=334$ with
respect to the mean density of the universe at $z$=0. During the
resampling we retained previous large-scale waves intact but included
additional small-scale waves, as described by \citet{klypin_etal01}.
The resampled Lagrangian region of each cluster was then re-simulated
with high dynamic range.

High-resolution simulations were run using 128$^3$ uniform grid and 8
levels of mesh refinement in computational boxes of $120h^{-1}$~Mpc
for CL1-CL3 and $80h^{-1}$~Mpc for CL4-CL11, which corresponds to the
dynamic range of $128\times 2^8=32768$ and the peak formal resolution
of $80/32,768\approx 2.44h^{-1}\ \rm kpc$, corresponding to the actual
resolution of $\approx 2\times 2.44\approx 5h^{-1}\ \rm kpc$. Only the
region of $\sim 3-10h^{-1}\ \rm Mpc$ around a cluster was adaptively
refined, the rest of the volume was followed on the uniform $128^3$
grid. The mass resolution corresponds to the effective $512^3$
particles in the entire box with a particle mass $m_p=1.07\times
10^{9}M_{\odot}$ and $3.16\times 10^{8}M_{\odot}$, or the Nyquist
wavelength of $\lambda_{\rm Ny}=0.469h^{-1}$ and $0.312h^{-1}$ {\it
comoving} megaparsec for CL1-3 and CL4-11, respectively, or
$0.018h^{-1}$ and $0.006h^{-1}$ Mpc in physical units at initial
redshift of the simulations. The dark matter particle mass in the
region around the cluster is therefore $(1-f_b)m_p=9.1\times
10^{8}h^{-1}{\rm\ M_{\odot}}$ for CL1-3 and $2.7\times
10^{8}h^{-1}{\rm\ M_{\odot}}$ for CL4-CL11 (where $f_b = \Omega_{\rm
b}/\Omega_{\rm m}$ = 0.1429), while other regions were simulated with
lower mass resolution.

As the zeroth-level fixed grid consisted of only $128^3$ cells, we
started the simulation already pre-refined to the 2nd level
($l=0,1,2$) in the high-resolution Lagrangian regions of clusters.
This is done to ensure that the cell size is equal to the mean
interparticle separation and all fluctuations present in the initial
conditions are evolved properly. During the simulation, refinements
were allowed to the maximum $l=8$ level and refinement criteria were
based on the local mass of DM and gas in each cell. The logic is to
keep the mass per cell approximately constant so that the refinements
are introduced to follow collapse of matter in a quasi-Lagrangian
fashion. For the DM, we refine the cell if it contains more than two
dark matter particles of the highest mass resolution specie.  For gas,
we allow the mesh refinement, if the cell contains gas mass larger
than four times DM particle mass scaled by the baryon fraction.  In
other words, the mesh is refined if the cell contains DM mass larger
than $2(1-f_b)m_p$ or gas mass larger than $=4f_bm_p$.  We analyze
clusters at the present-day epoch as well as their progenitors at
higher redshifts.

We repeated each cluster simulation with and without radiative cooling
and star formation.  The first set of ``adiabatic'' simulations have
included only the standard gasdynamics for the baryonic component
without radiative cooling and star formation.  The second set of
simulations included gasdynamics and several physical processes
critical to various aspects of galaxy formation: radiative cooling,
star formation, metal enrichment and thermal feedback due to the
supernovae type II and type Ia, self-consistent advection of metals,
metallicity dependent radiative cooling and UV heating due to
cosmological ionizing background \citep{haardt_madau96}. We will use
labels 'AD' and 'CSF' for the adiabatic simulations and the
simulations with gas cooling and star formation, respectively.

Cooling and heating rates take into account Compton heating and
cooling of plasma, UV heating, atomic and molecular cooling and are
tabulated for the temperature range $10^2<T<10^9$~K and a grid of
metallicities, and UV intensities using the {\tt Cloudy} code
\citep[ver. 96b4,][]{ferland_etal98}. The Cloudy cooling and heating
rates take into account metallicity of the gas, which is calculated
self-consistently in the simulation, so that the local cooling rates
depend on the local metallicity of the gas. Star formation in these
simulations was done using an observationally-motivated recipe
\citep[e.g.,][]{kennicutt98}: $\dot{\rho}_{\ast}=\rho_{\rm
gas}^{1.5}/t_{\ast}$, with $t_{\ast}=4\times 10^9$~yrs. The detailed
description and implementation of the star formation are provided in
Kravtsov et al. (2005).

The code also accounts for the stellar feedback on the surrounding
gas, including injection of energy and heavy elements (metals) via
stellar winds and supernovae and secular mass loss.  Once formed, each
stellar particle is treated as a single-age stellar population and its
feedback on the surrounding gas is implemented accordingly. More
specifically, in the simulations analyzed here, we assumed that a
stellar initial mass function (IMF) is described by the
\citet{miller_scalo79} functional form with stellar masses in the
range $0.1-100\ \rm M_{\odot}$. All stars more massive than
$M_{\ast}>8{\ \rm M_{\odot}}$ deposit $2\times 10^{51}$~ergs of
thermal energy in their parent cell\footnote{No delay of cooling was
introduced in these cells after SN energy release.} and a fraction
$f_{\rm Z}= {\rm min}(0.2,0.01M_{\ast}-0.06)$ of their mass as metals,
which crudely approximates the results of \citet{woosley_weaver95}. In
addition, stellar particles return a fraction of their mass and metals
to the surrounding gas at a secular rate $\dot{m}_{\rm
loss}=m_{\ast}\,\,C_0(t-t_{\rm birth} + T_0)^{-1}$ with $C_0=0.05$ and
$T_0=5$~Myr \citep{jungwiert_etal01}. The code also accounts for the
SNIa feedback assuming a rate that slowly increases with time and
broadly peaks at the population age of 1~Gyr. We assume that a
fraction of $1.5\times 10^{-2}$ of mass in stars between 3 and $8\ \rm
M_{\odot}$ explodes as SNIa over the entire population history and
each SNIa dumps $2\times 10^{51}$ ergs of thermal energy and ejects
$1.3\ \rm M_{\odot}$ of metals into the parent cell. For the assumed
IMF, 75 SNII (instantly) and 11 SNIa (over several billion years) are
produced by a $10^4\ \rm M_{\odot}$ stellar particle.

High-resolution\footnote{The mass and spatial resolution are high
enough to resolve galactic-size halos whose mass is as small as
$10^{-4}$ of the host cluster mass \citep{nagai_etal05}.}  and
inclusion of various physical processes are critical for assessing the
effects of galaxy formation on properties and evolution of the ICM.
For example, \citet{kravtsov_etal05} showed that gas cooling and star
formation significantly suppress gas mass fractions in clusters.
Interestingly, the results of numerical simulations agree quite well
with recent X-ray and SZE measurements \citep[e.g.,][Kravtsov et al.,
in preparation]{laroque_etal06}. Note also that the level of entropy
and metal abundance in these simulations also compare well with recent
X-ray observations.  Because the SZE signal depends linearly on the
gas mass fraction, these results indicate that the impact of galaxy
formation on the SZE flux may also be significant.

Throughout this paper we use estimates of $Y^{\rm int}$, mass and
other cluster observables within different commonly used radii,
defined by the total matter overdensity they enclose. We will use
radii $r_{2500}$, $r_{500}$, $r_{200}$ enclosing overdensities of
$\Delta_c=2500$, $500$, and $200$ with respect to the critical
density, $\rho_{\rm crit}$, as well as radii $r_{180}$ and $r_{\rm
vir}$ enclosing overdensities of $\Delta_m=180$ and $\Delta_{\rm vir}$
with respect to the mean density of the universe. The latter is equal
to $\Delta_{\rm vir}\approx 334$ at $z$=0 and $\approx 200$ at $z$=1
for the cosmology adopted in our simulations.  The virial radius and
masses of clusters within $r_{500c}$ for the CSF run at $z$=0 are
given in Table~\ref{tab:sim}.  For reference, we also give a spectral
X-ray temperature, $\langle T_{\rm spec} \rangle$, of individual
clusters extracted from mock {\it Chandra} analysis (Nagai, Vikhlinin
\& Kravtsov 2006, in preparation), and it is equivalent to a
single-temperature fit to the cluster spectrum extracted from the
radial range $0.15<r/r_{500c}<1.0$.

\begin{deluxetable}{lccccc}
%\onehalfspacing
\tablewidth{0pt}
\tablecolumns{6}
\tablecaption{Simulated cluster sample of the CSF run at $z$=0}
\tablehead{
\\
\multicolumn{1}{c}{Name}&
\multicolumn{1}{c}{}&
\multicolumn{1}{c}{$R_{\rm 500c}$} &
\multicolumn{1}{c}{$M_{\rm 500c}^{gas}$}  &
\multicolumn{1}{c}{$M_{\rm 500c}^{tot}$} &
\multicolumn{1}{c}{$\langle T_{\rm spec}\rangle$} 
\\ \\
\multicolumn{2}{c}{}&
\multicolumn{1}{c}{$h^{-1}$Mpc}&
\multicolumn{1}{c}{$h^{-1}10^{13}M_{\odot}$}& 
\multicolumn{1}{c}{$h^{-1}10^{14}M_{\odot}$} &
\multicolumn{1}{c}{keV} 
\\
}
\startdata
\\
CL1  & & 1.160 & 8.19 & 9.08 & 10.0 \\
CL2  & & 0.976 & 5.17 & 5.39 & 8.1  \\
CL3  & & 0.711 & 1.92 & 2.09 & 4.1  \\
CL4  & & 0.609 & 1.06 & 1.31 & 2.7  \\
CL5  & & 0.661 & 1.38 & 1.68 & 3.9  \\
CL6  & & 0.624 & 1.22 & 1.41 & 3.2  \\
CL7  & & 0.522 & 0.74 & 0.82 & 1.7  \\
CL8  & & 0.487 & 0.43 & 0.67 & 2.2  \\
CL9  & & 0.537 & 0.78 & 0.90 & 2.2  \\
CL10 & & 0.509 & 0.62 & 0.77 & 2.1  \\
CL11 & & 0.391 & 0.27 & 0.35 & 1.2  \\
\enddata
\label{tab:sim}
\end{deluxetable}
\vspace{0.8cm}

%-------------------
\section{Results}
\label{sec:results}
%-------------------

In this section, we study the impact of galaxy formation on the SZE
scaling relations using cosmological simulations of cluster formation.
Our general strategy is to assess their effects on cluster observables
and scaling relations by comparing two sets of simulations with and
without processes important for galaxy formation.

% Fig.1: Pressure profiles at z=0 
\begin{figure*}[t]  
  \vspace{-0.0cm}
  \hspace{0.0cm}
  \centerline{ \psfig{figure=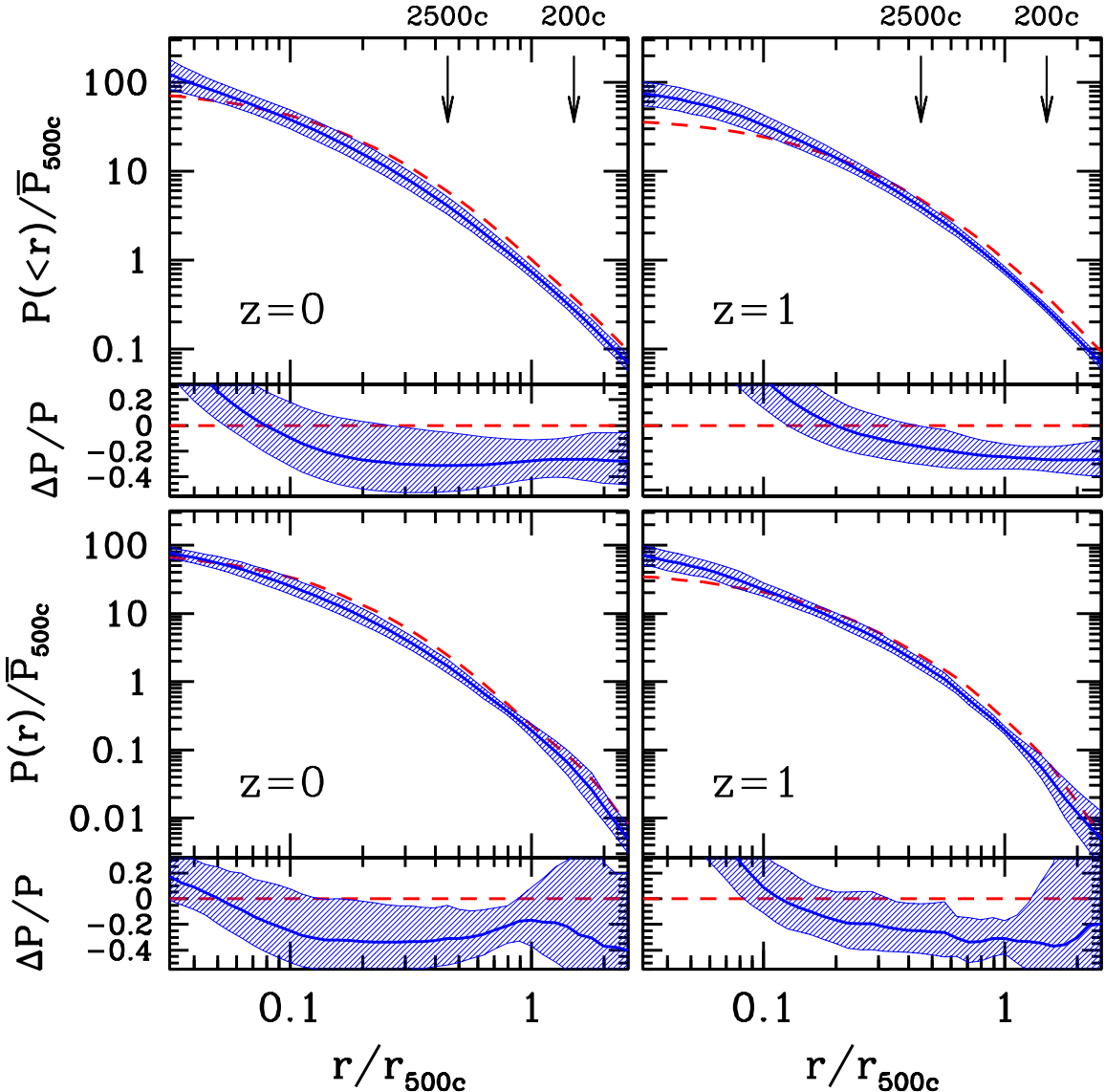,height=17cm} }
  \vspace{-1.5cm}
\caption{Cumulative ({\it top panels}) and differential ({\it
bottom panels}) pressure profiles for the eleven clusters used in our
analysis at $z$=0 ({\it left column}) and $z$=1 ({\it right column}).
The {\it dashed} and {\it solid} lines show mean profiles in the AD
and CSF simulations averaged over all clusters, respectively.  Note
that all profiles are normalized to the mean cumulative pressure at
$r_{500c}$ in the AD runs, $\bar{P}_{500c}$, at each epoch.  The {\it
bottom panel} of each figure shows the fractional deviation of the
mean profile of the CSF run relative to the mean profile of the AD
run.  In all panels, the {\it shaded} bands indicate 1$\sigma$ {\it
rms} scatter around the mean.  The vertical arrows in the top panels
indicate the radii enclosing overdensities of 2500 and 200 with
respect to the critical density at each epoch.}
\label{fig:profiles}
\end{figure*}

% Fig.2: 
\begin{figure}[t]  
  \vspace{-1.2cm}
  \hspace{3.2cm}
  \centerline{ \psfig{figure=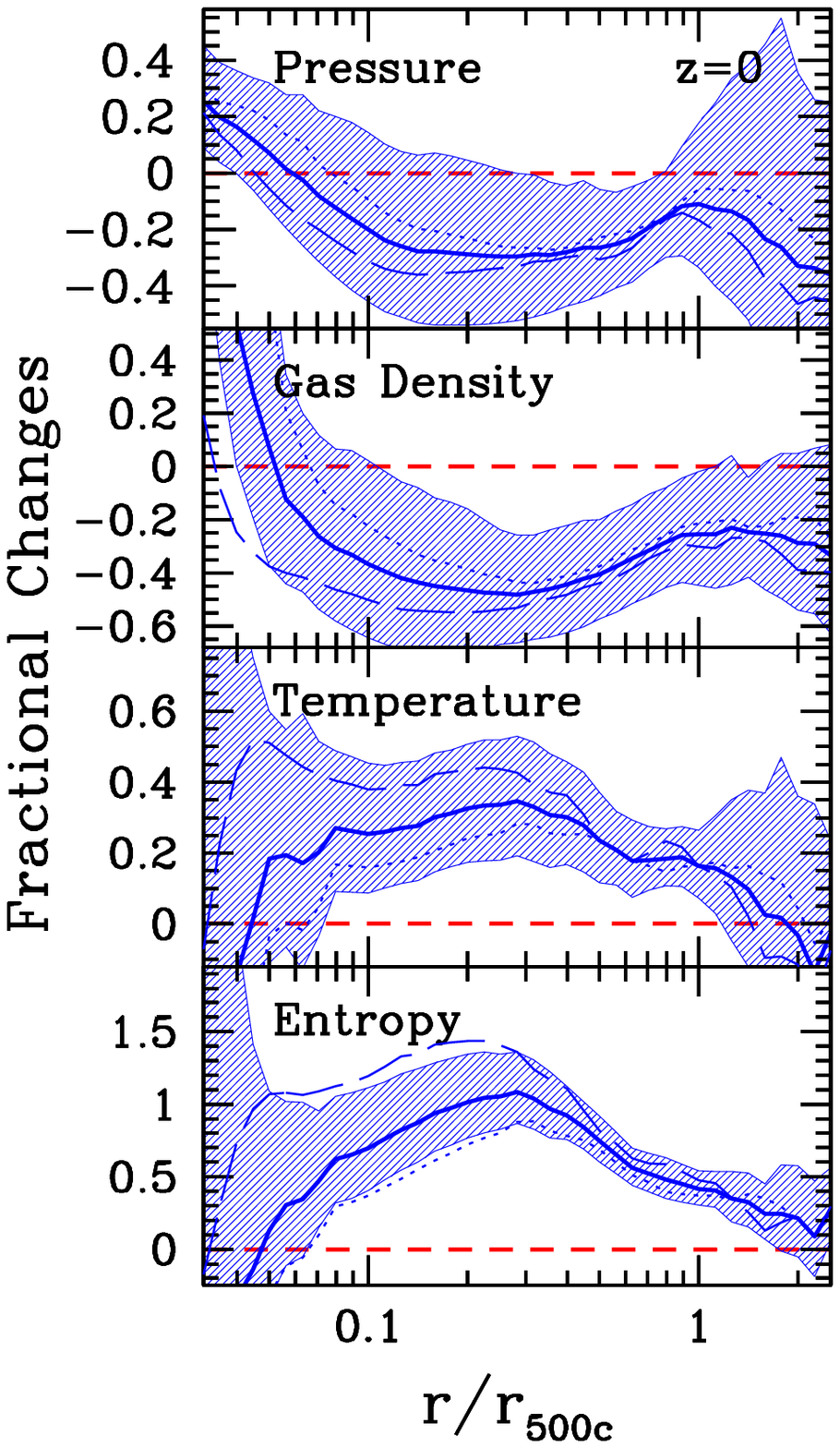,height=17cm} }
  \vspace{-1.5cm}
\caption{The impact of galaxy formation on pressure,
gas density, temperature and entropy profiles ({\it from top to
bottom}) of the simulated clusters at $z$=0.  The {\it solid} lines
show fractional changes in the mean differential radial profiles in
the CSF runs averaged over eleven clusters relative to the mean
profiles in the AD runs, indicated by lines at zero.  The {\it dotted}
and {\it dashed} lines show the same for subsamples of six most and
five least massive clusters, respectively.  Note that the mean profile
of each quantity is computed by first normalizing a profile of each
cluster by the value of the cumulative profile at $r_{500c}$ and
averaging the normalized profile over all
clusters. } \label{fig:profiles_comp}
\end{figure}

%-----------------------------
\subsection{Pressure Profiles} 
\label{sec:profiles}
%-----------------------------

Since the thermal SZE is sensitive to the pressure of the ICM, we will
start with the analysis of pressure profiles of
clusters. Figure~\ref{fig:profiles} shows a comparison of the mean
pressure profiles of clusters simulated in the adiabatic regime and
those with gas cooling and star formation at $z$=0 and $z$=1.  To
compute the mean profiles, we first normalize differential and
cumulative pressure profiles of individual clusters using the mean
integrated pressure within a sphere of $r_{500c}$, denoted as
$\bar{P}_{500c}$, and plot them as a function of the cluster-centric
radius in units of $r_{500c}$.  The mean profiles are computed by
averaging the normalized profiles over the entire cluster sample.  To
highlight the impact of gas cooling and star formation on the pressure
profiles, we normalize the pressure profile of the CSF run using
$\bar{P}_{500c}$ of the AD run for each cluster.  The shaded bands
show $1\sigma$ rms scatter around the mean profile of the CSF runs.
The mean and scatter of the profiles are computed for a logarithm of
pressure.

Fig.~\ref{fig:profiles} illustrates that clusters ranging in an order
of magnitude in mass exhibit remarkably similar pressure profiles at
all radii.  The differential pressure profile declines by nearly four
orders of magnitude from the center to $r_{500c}$, and the scatter is
small ($\approx 20\%$) throughout clusters.  In the outskirts, the
radial profile of gas pressure generally falls faster than that of gas
or dark matter density.  This is because pressure is a product of gas
density and gas mass-weighted temperature for the ideal gas, and the
latter also declines with radius
\citep[see][]{motl_etal05,vikhlinin_etal05c}.  For example, a typical
power-law slope of the differential pressure profile is $\approx -3.5$
at $r_{500c}$, which is steeper than a typical slope of the density
profile of $\approx -3$ seen in cosmological simulations
\citep{navarro_etal96,navarro_etal97} as well as recent X-ray
observations \citep{vikhlinin_etal05b}.

One of the main results from this analysis is that gas pressure is
suppressed in a region outside $r \gtrsim 0.1 r_{500c}$ in the CSF
runs compared to the AD runs.  The cumulative pressure at $r_{500c}$,
for example, is suppressed by $\approx 25\%$ on average, and this
effect is relatively constant ($\approx 25-35\%$) in a region outside
$>0.2r_{500c}$ at $z$=0.  In the center, gas pressure is significantly
more concentrated, because of strong gas cooling and contraction of
dark matter \citep{gnedin_etal04}.  Note, however, that this high
pressured cluster core contributes very little to the cumulative
pressure at large radii.  For example, the inner regions within
$<0.1r_{500c}$ contribute less than 2\% of the total SZE signal within
$r_{500c}$. Thus, the SZE signal integrated out to a sufficiently
large radius is insensitive to properties of cluster cores.

Although general trends are similar, the suppression of gas pressure
varies more strongly with radius at $z$=1, because a dense, cool core
dominates a larger fraction of $r_{500c}$ at high
redshift. Nevertheless, we find that the suppression of the cumulative
pressure at $r_{500c}$ is comparable at $z$=0 and 1.  However, the
difference becomes more apparent in the inner region; for example, the
effect at $r_{2500c}$ changes from 35\% at $z$=0 to 20\% at $z$=1.  If
a similar trend exists in real clusters this implies that redshift
evolution of the SZE scaling relations is expected to deviate from the
prediction of the self-similar model, if the SZE flux is computed
within the aperture considerably smaller than $r_{500c}$.

In order to better understand the suppression of gas pressure, we
examine the impact of gas cooling and star formation on density,
temperature and entropy profiles of the ICM.
Figure~\ref{fig:profiles_comp} shows that the suppression of pressure
is accompanied by strong suppression in gas density and moderate
increase in gas mass-weighted temperature.  An entropy profile
provides further insights into these phenomena, because it uniquely
characterizes thermodynamic properties of the ICM, given a shape of
its confining potential well (see Voit et al. 2002 and references
therein).  Fig.~\ref{fig:profiles_comp} shows that the entropy of the
ICM, defined as $S\equiv T/n^{2/3}_e$, is enhanced from $0.05\times
r_{500c}$ out to $2\times r_{500c}$.  The effect peaks around
$0.3r_{500c}$, and the entropy is enhanced by about 40\% at
$r_{500c}$. In our simulations, gas cooling and star formation can
increase the entropy of the gas in one of two ways: (1) the level of
entropy is elevated throughout as gas cooling and star formation
proceed in the cluster center \citep{voit_bryan01}\footnote{Gas
cooling and star formation remove low entropy gas and cause an inflow
of high entropy gas from outside to maintain the stability of the
cluster core.  In this way, the low-entropy gas in the inner region is
gradually replenished with the high-entropy gas flowing in from
outside, and the level of entropy is gradually elevated throughout the
cluster as gas cooling and star formation proceed in the cluster
core.} or (2) direct heating of the ICM by energy injection from
supernova explosions.  The former is likely a dominant mechanism that
shapes the overall entropy distribution, while the second process is
important for regulating a rate of gas cooling in high density regions
(e.g., a vicinity of a central galaxy and cluster galaxies), where
stars are forming with sufficiently high rates.

To examine a mass dependence of the effects, we split the sample in
half and show the mean profiles of six most massive ($T_X\gtrsim
2$keV) and five least massive ($T_X\lesssim 2$keV) clusters in
Fig.~\ref{fig:profiles_comp}. Note that the former includes clusters
dominated by bremsstrahlung or those that are about equal in the
importance of bremsstrahlung and line cooling, while the latter
includes clusters dominated by line
emission. Fig.~\ref{fig:profiles_comp} shows that a shape and
normalization of the pressure profiles exhibit the smallest systematic
trend with mass.  In the outskirts ($r>0.3r_{500c}$) of the clusters,
there is a similarly small systematic trend with mass for gas density,
temperature and entropy profiles.  In the inner regions, on the other
hand, we find a slightly more pronounced mass dependence in all
profiles.  Among them, the pressure profiles show the smallest trend
with mass.  In less massive clusters, changes in gas density and
temperature become larger but with opposite signs; therefore, these
effects cancel each other to give the smallest mass dependence on
pressure.  In contrast, the effects add constructively to give the
largest mass dependence on entropy.

The main conclusions from the profile analysis are that gas cooling
and star formation suppress the amplitude of gas pressure and
therefore the SZE signal, and the magnitude of the effects depends
rather weakly on the cluster mass.  These results suggest that
processes of galaxy formation modify normalization of the SZE
flux-mass relation, but its impact on slope is expected to be small.
This is a main topic of the discussion in the next section.

% Fig 3: SZ flux-Mtot relation at z=0 & z=1 
\begin{figure*}[t]  
  \vspace{-0.0cm} \hspace{-0.0cm} \centerline{
  \psfig{figure=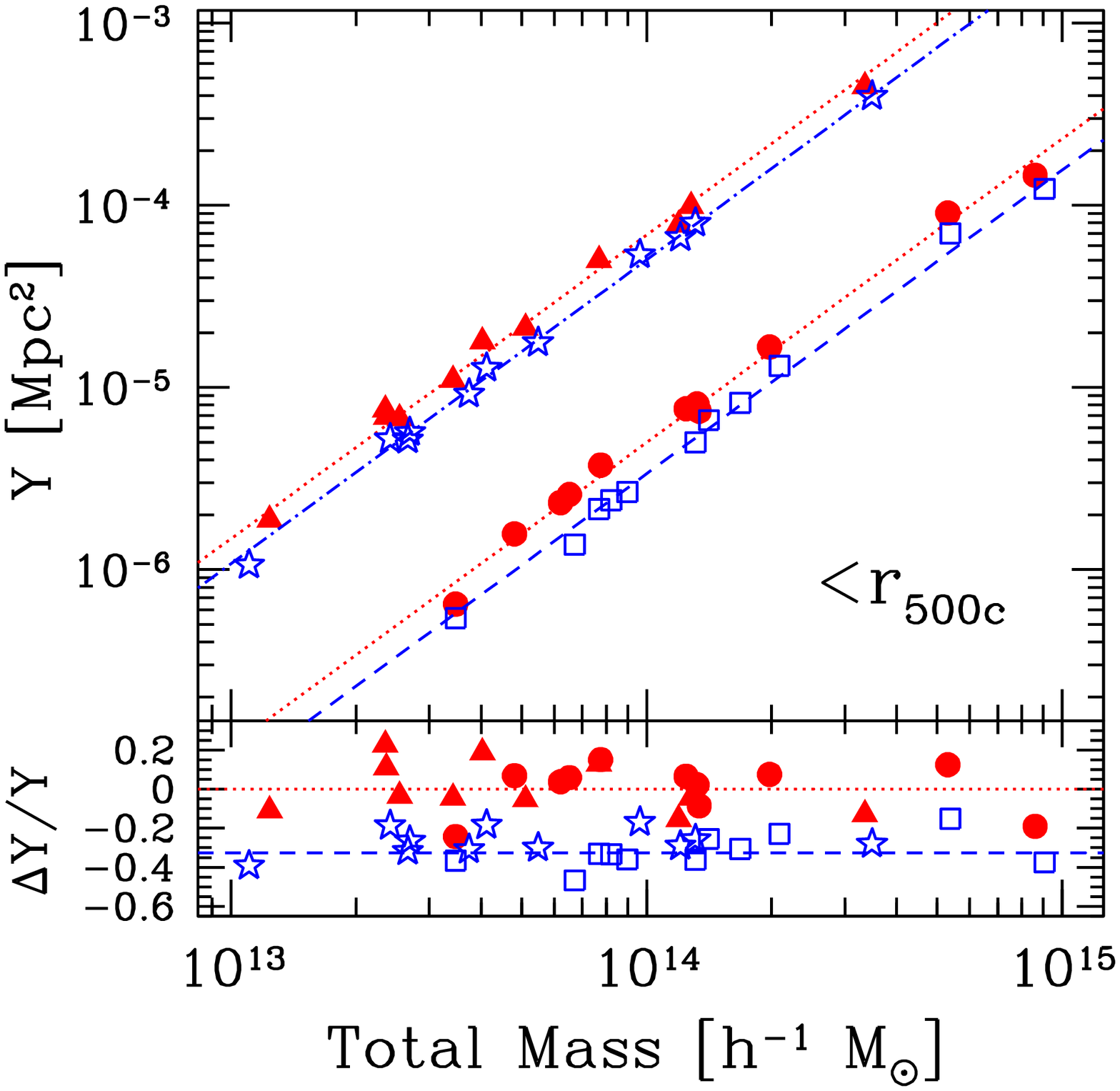,height=8.5cm}
  \psfig{figure=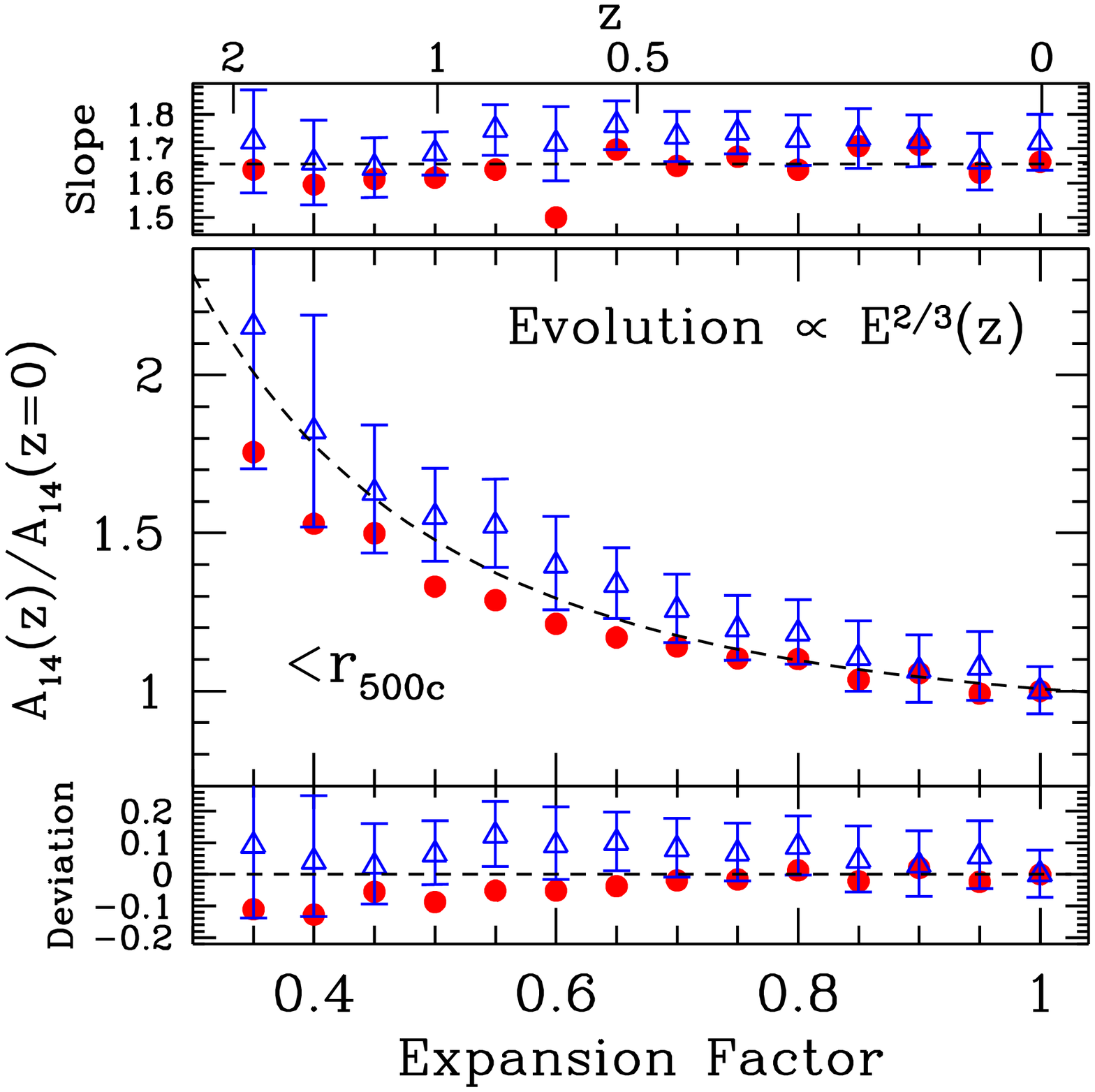,height=8.5cm} } \vspace{-0.0cm}
\caption{The SZE flux-total mass relations for eleven clusters in
the AD ({\it solid symbols}) and CSF ({\it open symbols})
simulations. {\it Left-top panel:} the relation between the integrated
compton-Y vs. total mass enclosed within a sphere of radius $r_{500c}$
at $z$=0 ({\it bottom lines and points}) and $z$=1 ({\it top lines and
points}).  For clarity, the $z$=1 relations are shifted upward by a
factor of ten.  The {\it solid circles} and {\it solid triangles}
indicate the AD runs at $z$=0 and 1, and the {\it open squares} and
{\it open stars} are the CSF runs at $z=0 $ and 1, respectively.  At
each epoch, the {\it dotted} and {\it dashed} lines show the best-fit
relations for the AD and CSF runs with a slope fixed at the
self-similar slope of 5/3.  {\it Left-bottom panel:} the fractional
deviation of individual clusters from the best-fit relation of the AD
runs at each epoch. The {\it dashed} line indicates the best-fit
relation of the CSF runs at $z$=0. {\it Right panel:} the redshift
evolution of slope ({\it top}) and normalization ({\it middle}) of the
SZE flux-total mass relation between $z$=2 and the present day.  The
{\it solid circles} and {\it open squares} indicate the AD and CSF
runs, respectively. The {\it dashed} lines indicate the slope of 5/3
({\it top}) and the evolution in normalization predicted by the
self-similar model ({\it middle}).  The errorbars indicate 2$\sigma$
confidence region of the best-fit slope and normalization in the CSF
runs at each epoch.  The {\it bottom} panel shows a fractional
deviation of the best-fit normalization from the self-similar
prediction as a function of time.}
\label{fig:szm_r500c}
\end{figure*}

%---------------------------------------
\subsection{The SZE scaling relations}
\label{sec:SZscaling}
%---------------------------------------

The SZE flux-total mass (SZE-M) relation is a relation most directly
relevant for cosmological application.  Figure~\ref{fig:szm_r500c}
shows a comparison of the SZE-total mass scaling relation at
$r_{500c}$ and its redshift evolution for a sample of eleven clusters
in the AD and CSF runs.  At each epoch, we performed fits to the
sample of simulated clusters using the simple power-law relation,
\begin{equation}
Y = A_{14}\times10^{-6} \left( \frac{M}{10^{14}h^{-1}M_{\odot}}
\right)^{\alpha_M},
\end{equation}
where $A_{14}$ is the normalization at $10^{14}h^{-1}M_{\odot}$ in
units of $10^{-6}$ and $\alpha_M$ is the slope. In practice, we fit a
straight line to the log(Y)-log(M) relation by minimizing $\chi^2$.
Table~\ref{tab:SZmtot}-\ref{tab:SZtm} lists the best-fit normalization
and slope measured at different radii, including $r_{180m}$, $r_{\rm
vir}$, $r_{200c}$, $r_{500c}$ and $r_{2500c}$.  The values in the
adiabatic simulations are marked 'ad', while those in the simulations
with gas cooling and star formation are marked 'csf' (e.g.,
$A_{14}^{{\rm csf}}$).  The best-fit parameters in these tables and
the right panel of Fig.~\ref{fig:szm_r500c} are obtained by fitting
for both normalization and slope simultaneously.  To highlight a
deviation from the self-similar slope, the best-fit relations shown in
the left panel of Fig.~\ref{fig:szm_r500c} are obtained by fixing the
slope to the predicated value of 5/3. Note that fitting for the slope
changes the best-fit normalization by no more than 5\% at both $z$=0
and $z$=1.

\begin{figure}[t]  
  \hspace{-0.0cm}
  \vspace{-0.6cm}
  \centerline{ \psfig{figure=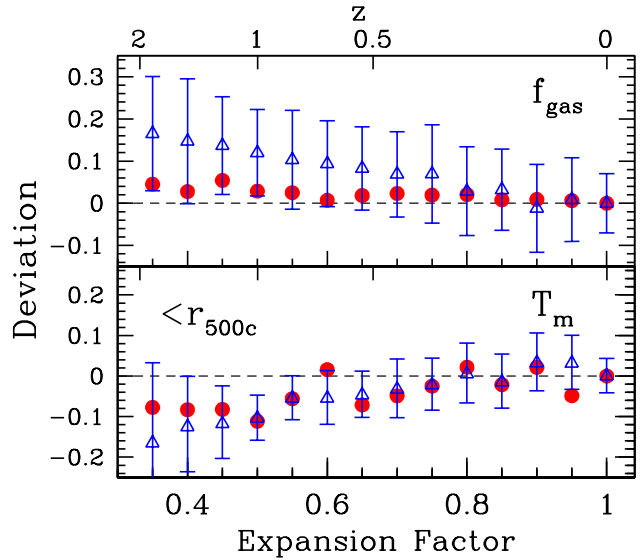,height=9.2cm} }
  \vspace{-0.5cm}
\caption{The evolution of gas mass
fraction ({\it top}) and mass-weighted temperature ({\it bottom}) at
$M_{500c}=10^{14}h^{-1}M_{\odot}$.  The fractional deviation of the
best-fit normalization from the self-similar prediction, indicated by
the {\it dashed} line.  The errorbars indicate 2$\sigma$ statistical
uncertainties in the CSF runs at each
epoch.}
\label{fig:change_evol_r500c}
\end{figure}

\begin{figure*}[t]  
  \vspace{-0.0cm}
  \hspace{-0.0cm}
  \centerline{
    \psfig{figure=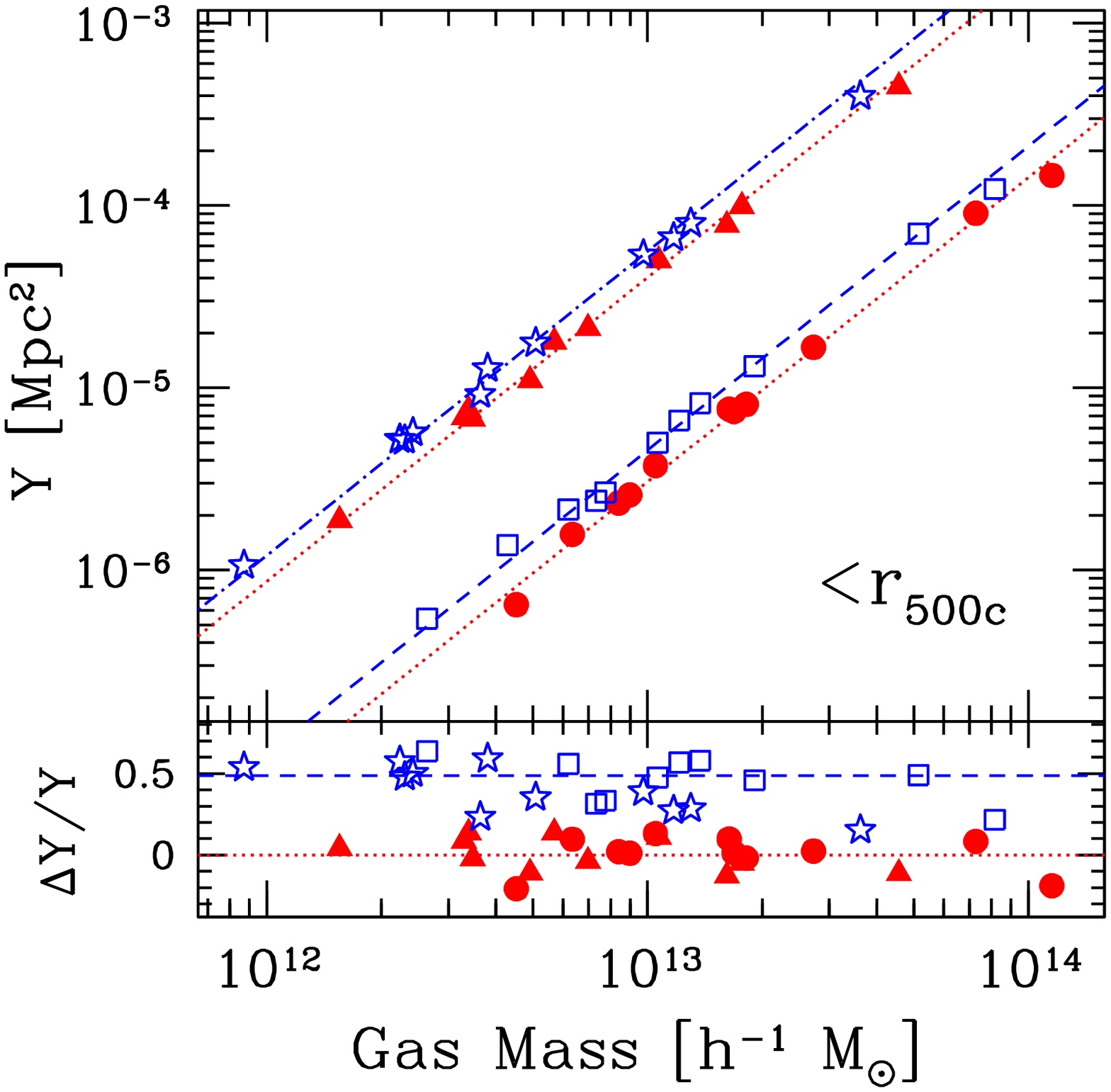,height=7.5cm}
    \psfig{figure=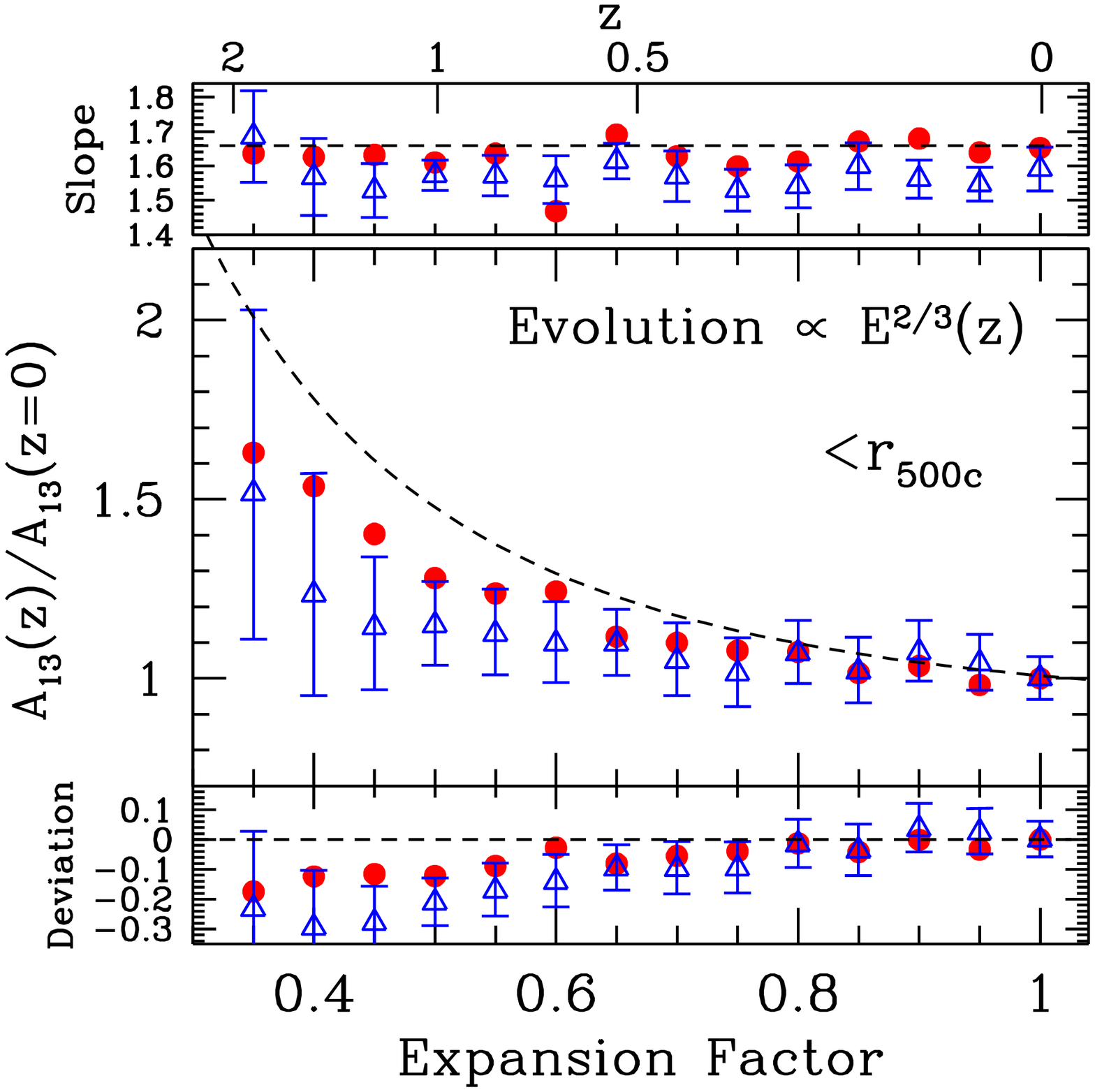,height=7.5cm}
  }
  \centerline{
    \psfig{figure=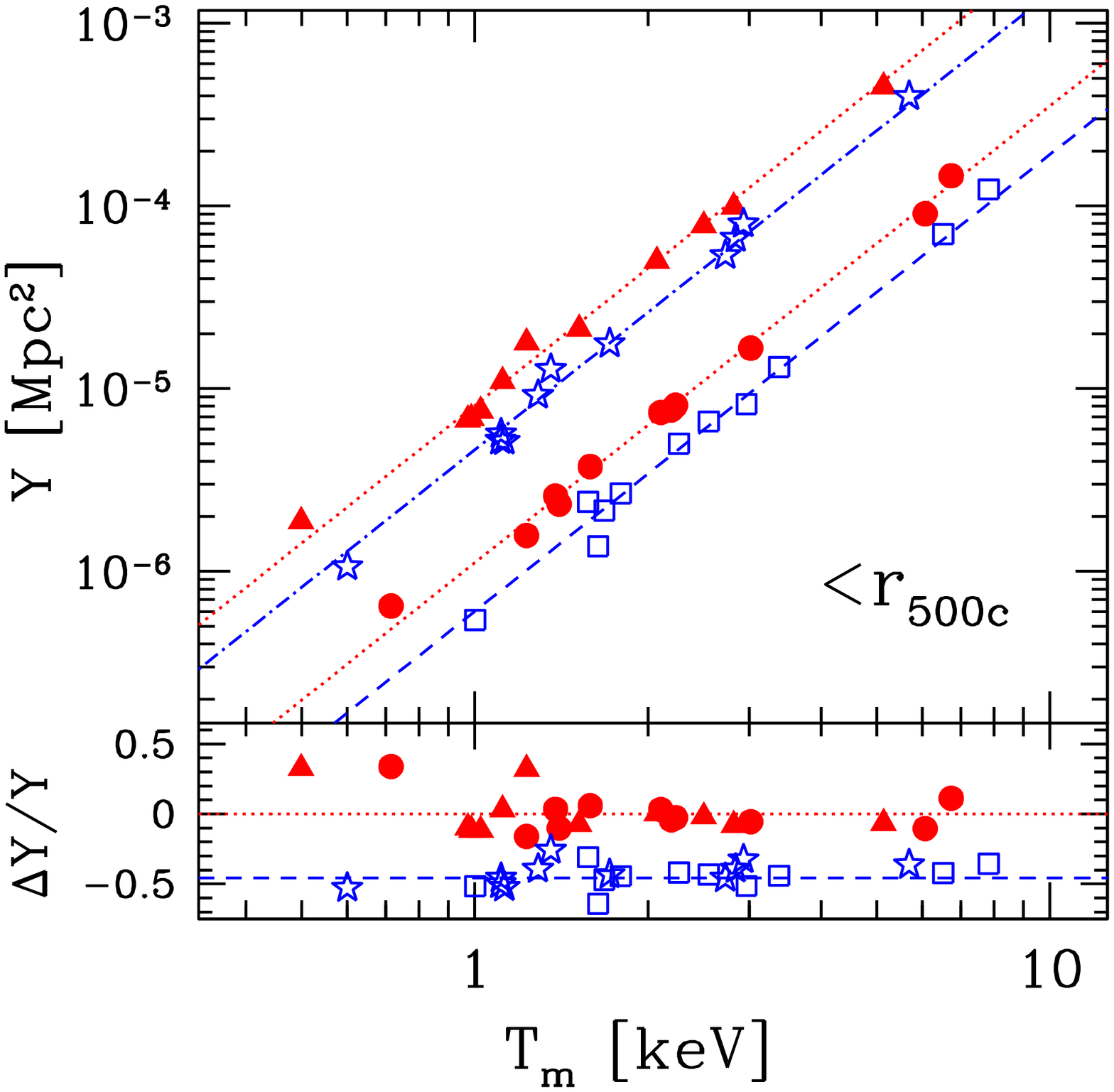,height=7.5cm}
    \psfig{figure=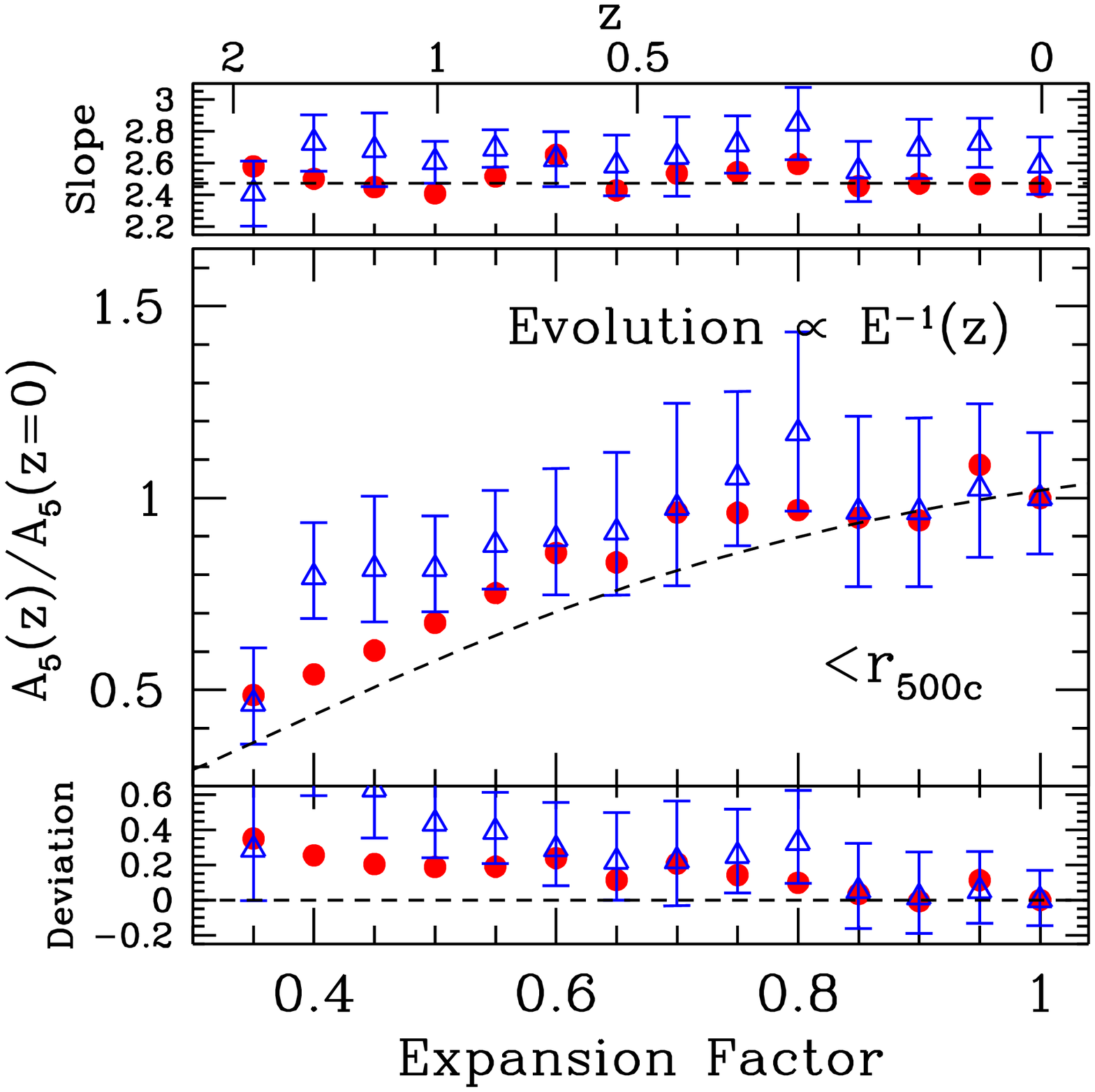,height=7.5cm}
  }
  \vspace{-0.2cm}
\caption{ {\it Top panels:} the SZE flux-gas mass relations.  The {\it
left panel} shows the scaling relations at $z$=0 ({\it bottom lines
and points}) and 1 ({\it top lines and points}).  The {\it right
panel} shows the redshift evolution in slope and normalization between
$z$=2 and 0.  The errorbars indicate 2$\sigma$ confidence region of
the best-fit slope and normalization in the CSF runs at each epoch.
Point and line types are the same as in Fig.~\ref{fig:szm_r500c}. {\it
Bottom panels:} the relation between the SZE flux and the
mass-weighted temperature.} 
\label{fig:szmg_r500c} 
\end{figure*}

In the right panel of Fig.~\ref{fig:szm_r500c}, errorbars indicate
2$\sigma$ confidence region of the best-fit slope and normalization in
the CSF simulations at each epoch. Since the redshift evolution of
normalization is measured relative to the relation at $z$=0, errorbars
at $z>$0 include uncertainties in the best-fit normalizations at each
epoch and $z$=0 added in quadrature.  Note also that the size of the
errorbars are comparable for the AD simulations.  Errorbars shown in
Fig.~\ref{fig:change_evol_r500c} and Fig.~\ref{fig:szmg_r500c} are
also computed in the same way.

The left panel shows that the SZE signal integrated within a sphere of
$r_{500c}$ correlates very strongly with the enclosed cluster mass at
both $z$=0 and 1.  The {\it rms} scatter is about 10-15\% in both AD
and CSF runs.  Moreover, we find that the slope of the AD run is in a
very good agreement with the predicted slope of the self-similar
model.  The best-fit slope of the CSF run is also not very different
from the predicted slope of 5/3, but there is an indication that the
slope may be systematically steeper by $\lesssim 0.1$.  Unfortunately,
however, the current sample size is too small to rule out the
prediction of the self-similar model with any statistical
significance.  A larger sample of simulated clusters is clearly needed
to assess this effect in future studies.

While the tightness and slope of the relations are relatively
unaffected, gas cooling and star formation significantly modify
normalization of the SZE-total mass relation.  Compared to the AD run,
the normalization of the CSF run is lower by 41, 34 and 27\% at
$r_{2500c}$, $r_{500c}$ and $r_{180m}$.

The right panel shows that the redshift evolution of the SZE-total
mass relation is consistent with the predictions of the self-similar
model in the entire redshift range considered here. First, the slope
is constant with redshift.  Second, the redshift evolution of the
normalization is in a good agreement with the predicted evolution.  We
also find similar results at other radii considered in this paper.

% Fig.4: Change of fgas & Tm at z=0 and z=1 
\begin{figure}[t]  
  \vspace{-0.0cm}
  \hspace{-0.0cm}
  \vspace{-0.6cm}
  \centerline{ \psfig{figure=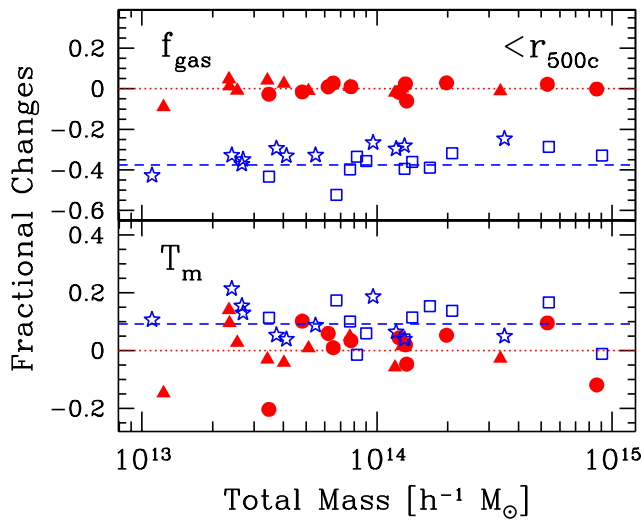,height=9.2cm} }
  \vspace{-0.5cm}
\caption{Fractional changes in gas
mass fraction ({\it top}) and mass-weighted temperature ({\it bottom})
enclosed within a sphere of radius $r_{500c}$ between the AD and CSF
simulations as a function of cluster mass. The {\it solid circles} and
{\it solid triangles} are the AD runs at $z$=0 and 1, and the {\it
open squares} and {\it open stars} are the CSF runs at these epochs,
respectively.  The {\it dashed} line indicates a mean fractional
change in $f_{gas}$ or $T_m$ in the CSF run at z=0 compared to that of
the AD run, indicated by the {\it dotted line} at
zero.}
\label{fig:change_r500c}
\end{figure}

% Fig.5:  Effects of CSF on the mass profiles
\begin{figure}[t]  
  \hspace{-0.0cm}
  \vspace{-0.6cm}
  \centerline{ \psfig{figure=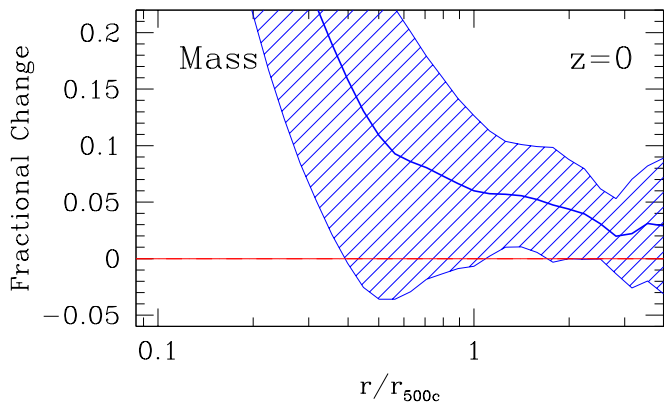,height=9.2cm} }
  \vspace{-2.8cm}
\caption{ Effects of gas cooling and star formation on total mass
profiles of simulated clusters at z=0. The {\it solid} line shows a
change in the total mass profiles in the CSF simulations compared to
that in the AD simulations averaged over {\it ten} clusters. For each
cluster, we compute a fractional change in the total mass profiles of
the CSF and AD runs, $(M_{\rm CSF}-M_{\rm A})/M_{\rm A}$, at the same
physical radius, where $M_{\rm CSF}$ and $M_{\rm A}$ are the total
mass profiles of the CSF and AD runs.  We then plot the fractional
change between two runs as a function of radius in units of $r_{500c}$
and average over ten clusters, while excluding one cluster
experiencing a major merger at z=0.}
\label{fig:massprof} 
\end{figure}

% Fig.6: Redshift evolution of the changes in fgas & Tm 

To better understand the impact of galaxy formation on the SZE
observable-mass relation, we also examine their effects on $f_{gas}$
and $T_{m}$ in Figure~\ref{fig:change_r500c}. Compared to the AD run,
$f_{gas}$ is lower by 38\% in the CSF run; however, its effect on the
SZE flux is offset partially by an increase in $T_{m}$ by 9\%. The net
effect is a reduction of the SZE signal by 32\%, which falls slightly
short of a 34\% change in the normalization of the SZE-M relation.
This illustrates that changes in $f_{gas}$ and $T_{m}$ alone do not
fully account for the change in normalization.

In addition, gas cooling and star formation slightly modify the total
cluster mass.  Figure~\ref{fig:massprof} illustrates that inclusion of
these processes cause an increase in $M_{tot}$ at $r_{500c}$ by about
6\% on average.  The effect becomes larger at a smaller radii, and it
is about 10\% at $r_{2500c}$.  This effect causes shifts in the SZE-M
relation in both x- and y-directions by the same amount.  However,
since the slope of the relation is $\approx$5/3, the 6\% shift in the
positive x-axis translates into a 8\% shift in the negative y-axis.
The net effect is therefore a negative 3\% change in normalization due
to the change in $M_{tot}$.  Putting them all together, we find that
the 34\% decrease in normalization of the SZE-M relation at $r_{500c}$
is due to combined effects of the 38\% decrease in $f_{gas}$, the 9\%
increase in $T_m$ and the net 3\% decrease in normalization due to the
change in $M_{tot}$.  Note also that inclusion of gas cooling and star
formation lowers the integrated SZE signal ($Y^{int}$) and the gas
mass ($M_g$) within $r_{500c}$ by $29\%$ and $35\%$, respectively.

% Fig.7: SZE flux-Mgas relations

Finally, we check the redshift evolution of $f_{gas}$ and $T_m$.  The
self-similar model predicts that $f_{gas}$ is constant with time,
while $T_m$ of a given clutter mass evolves with time according to
Eq.~\ref{eq:mt}.  To examine deviations from the predicted evolution,
Figure~\ref{fig:change_evol_r500c} plots the deviations of the
best-fit normalization of the $f_{gas}-M$ and the $T_m-M$ relations at
$M_{500c}=10^{14}h^{-1}M_{\odot}$ from the self-similar prediction as
a function of redshift for the AD and CSF runs.
Fig.~\ref{fig:change_evol_r500c} shows that $f_{gas}$ of a given mass
in the AD runs is consistent with the predicted evolution of the
self-similar model, while gas cooling and star formation cause
deviations, which increase toward higher redshifts.  In the CSF runs,
$f_{gas}$ is higher by about 12\% at $z$=1 and 17\% at $z$=2 compared
to their values at $z$=0.  $T_{m}$, on the other hand, is lower by an
almost equal amount at higher redshifts in both the AD and CSF runs.
Since the SZE signal is linearly proportional to both of these
quantities, the evolution in $f_{gas}$ is canceled almost exactly by
the evolution in $T_m$. Note also that we did not find any systematic
evolution in the impact of gas cooling and star formation on the total
mass with redshift.  This explains why the SZE signal of a given mass
shows very little evolution beyond the evolution predicted by the
self-similar model.

So far, we have focused on the SZE-total mass relation because of its
direct relevance for cosmological studies.  Unfortunately, however,
the cluster mass is not directly observable, and making unbiased
measurements of the SZE-M relation is difficult in
practice. Therefore, it is also useful to study relations between the
SZE signal and cluster properties that can be measured more reliably
from observations, as they can provide useful direct tests of
simulations.

Here, we consider two such quantities: gas mass ($M_{g}$) and
mass-weighted temperature ($T_m$) of clusters.
Figure~\ref{fig:szmg_r500c} shows SZE-$M_g$ and SZE-$T_m$ relations at
$r_{500c}$ and evolution of their slope and normalization with
redshift.  At each timestep, we perform a fit to the entire sample of
simulated clusters using the following simple power-law relations,
\begin{eqnarray}
Y & = & A_{13}\times10^{-6} \left( \frac{M_g}{10^{13}h^{-1}M_{\odot}} \right)^{\alpha_{M_g}} \\
Y & = & A_{5}\times10^{-5} \left( \frac{T_m}{5{\rm keV}} \right)^{\alpha_T}
\end{eqnarray}
where $A_{13}$ is the normalization at $10^{13}h^{-1}M_{\odot}$ in
units of $10^{-6}$ and $\alpha_{M_g}$ is the slope of the SZE-$M_g$
relation.  Similarly, $A_{5}$ is the normalization at 5~keV in units
of $10^{-5}$ and $\alpha_{T}$ is the slope of the SZE-$T_m$ relation.
Table~\ref{tab:SZmtot}-\ref{tab:SZtm} list the best-fit normalizations
and slopes measured at different radii for a $z$=0 sample.  The
errorbars indicate a 2$\sigma$ confidence region of the best-fit slope
and normalization in the CSF runs at each epoch.

Fig.~\ref{fig:szmg_r500c} shows that both the SZE-$M_g$ and the
SZE-$T_m$ relations are as equally tight as the SZE-M relation.  The
best-fit slopes of the AD runs are in a very good agreement with the
self-similar slope.  The best-fit slopes of the CSF runs, on the other
hand, are marginally consistent with the predicted slope, and there
are indications that the slopes in the CSF run may be slightly smaller
(larger) for the SZE-$M_g$(SZE-$T_m$) relations.  We also find that
the slope is constant with time in both the AD and CSF runs.

Existence of tight relations between the SZE observable and cluster
masses and temperature is encouraging news for future cosmological
studies.  In fact, the tightness of these relations is rather
remarkable, because the samples analyzed here include clusters in a
wide range of dynamical states.  This indicates that the integrated
thermal SZE signal is not very sensitive to dynamical states of
clusters, making the thermal SZE an excellent proxy of cluster mass.
The reason that the thermal SZE works so well is as follows.
Fundamentally, the ICM is trapped in the external potential of the
dark matter and the gas must remain in approximate hydrostatic
equilibrium in the dark matter potential provided that the
perturbations are small compared to the energy scale of the cluster
itself.  The pressure profile is therefore determined by approximate
force balance against gravity, and the thermal SZE measures the
integral of that profile.

While the tightness and slopes of these relations are relatively
unaffected, gas cooling and star formation have a large impact on
their normalization.  Fig.~\ref{fig:szmg_r500c} shows that the
normalization of the CSF run is higher by 49\% in the SZE-$M_g$
relation at $r_{500c}$, while it is lower by 39\% in the SZE-$T_m$
relation.  Note that the change in the normalization of the SZE-$M_g$
relation has an opposite sign from the change in the SZE-M relation,
because gas cooling and star formation significantly reduce cluster
gas mass.  Recall that the inclusion of these physical processes
reduces the SZE signal by 29\% and $M_g$ by 35\% at $r_{500c}$.  For
the slope of 5/3, a decrease in $M_g$ by 35\% in the x-direction
corresponds to an increase in normalization by a factor of two.
Combined with a reduction of the SZE signal, the net effect is an
increase in normalization by 49\% at $r_{500c}$.  Note that the
effects become larger in the inner region of the clusters.  Similar
accounting works for the SZE-$T_m$ relation.

In contrast to the SZE-M relation, gas cooling and star formation
modify the redshift evolution of normalization in the SZE-$M_g$ and
the SZE-$T_m$ relations.  In the CSF run, the normalization of the
SZE-$M_g$ relation is systematically lower by $\approx$20\% at $z$=1
than the prediction of the self-similar evolution model, while the
SZE-$T_m$ relation is higher by $\approx$40\% at $z$=1.  The departure
from the predicted evolution is mainly due to the evolution in
$f_{gas}$ and $T_{m}$, not the SZE signal, as discussed above (see
also Fig.~\ref{fig:change_evol_r500c}).  Note also that the redshift
evolution of the AD simulations is consistent with the self-similar
model at all redshift.

%%%%%%%%%%%%%%%%%%%%%%%%%%%%%%%%%%%%%%%%%%%%%%%%%%%%%%%%%

\begin{deluxetable}{lccccccc}
%\rotate
%\onehalfspacing
\tablewidth{0pt}
\tablecolumns{8}
\tablecaption{Normalization and slope of the SZE-Total Mass relations at $z$=0}
\tablehead{
\\
\multicolumn{1}{c}{$\Delta$}&
\multicolumn{1}{c}{} &
\multicolumn{1}{c}{$A_{14}^{\rm ad}$} &
\multicolumn{1}{c}{$\alpha_M^{\rm ad}$} &
\multicolumn{1}{c}{} &
\multicolumn{1}{c}{$A_{14}^{\rm csf}$} &
\multicolumn{1}{c}{$\alpha_M^{\rm csf}$} 
}
\startdata
\\
180m  & & $2.59^{+0.31}_{-0.28}$ & 1.63$\pm$0.09 & & $1.88^{+0.18}_{-0.17}$ & 1.68$\pm$0.07 \\ 
vir   & & $3.11^{+0.38}_{-0.34}$ & 1.61$\pm$0.09 & & $2.20^{+0.22}_{-0.20}$ & 1.67$\pm$0.08 \\ 
200c  & & $3.78^{+0.42}_{-0.38}$ & 1.62$\pm$0.10 & & $2.56^{+0.24}_{-0.22}$ & 1.70$\pm$0.08 \\ 
500c  & & $4.99^{+0.46}_{-0.42}$ & 1.66$\pm$0.09 & & $3.29^{+0.26}_{-0.24}$ & 1.73$\pm$0.08 \\ 
2500c & & $8.82^{+0.65}_{-0.60}$ & 1.68$\pm$0.07 & & $5.22^{+0.45}_{-0.41}$ & 1.77$\pm$0.09 \\
\enddata
%\normalsize
\label{tab:SZmtot}
\end{deluxetable}
\vspace{0.8cm}

%%%%%%%%%%%%%%%%%%%%%%%%%%%%%%%%%%%%%%%%%%%%%%%%%%%%%%%%%

\begin{deluxetable}{lccccccc}
%\rotate
%\onehalfspacing
\tablewidth{0pt}
\tablecolumns{8}
\tablecaption{Normalization and slope of the SZE-Gas Mass relations at $z$=0}
\tablehead{
\\
\multicolumn{1}{c}{$\Delta$} &
\multicolumn{1}{c}{} &
\multicolumn{1}{c}{$A_{13}^{\rm ad}$} &
\multicolumn{1}{c}{$\alpha_{Mg}^{\rm ad}$} & 
\multicolumn{1}{c}{} &
\multicolumn{1}{c}{$A_{13}^{\rm csf}$} &
\multicolumn{1}{c}{$\alpha_{Mg}^{\rm csf}$} 
}
\startdata
\\
180m  & & $1.55^{+0.23}_{-0.20}$ & 1.63$\pm$0.09 & & $2.19^{+0.22}_{-0.20}$ & 1.60$\pm$0.07 \\ 
vir   & & $1.86^{+0.25}_{-0.22}$ & 1.61$\pm$0.09 & & $2.61^{+0.23}_{-0.21}$ & 1.60$\pm$0.07 \\ 
200c  & & $2.29^{+0.22}_{-0.20}$ & 1.62$\pm$0.07 & & $3.24^{+0.21}_{-0.20}$ & 1.60$\pm$0.06 \\ 
500c  & & $3.08^{+0.27}_{-0.25}$ & 1.65$\pm$0.08 & & $4.59^{+0.28}_{-0.27}$ & 1.60$\pm$0.06 \\ 
2500c & & $6.31^{+0.43}_{-0.41}$ & 1.64$\pm$0.07 & & $9.27^{+0.60}_{-0.56}$ & 1.58$\pm$0.05 \\ 
\enddata
%\normalsize
\label{tab:SZmgas}
\end{deluxetable}
\vspace{0.8cm}

%%%%%%%%%%%%%%%%%%%%%%%%%%%%%%%%%%%%%%%%%%%%%%%%%%%%%%%%%

\begin{deluxetable}{lccccccc}
%\rotate
%\onehalfspacing
\tablewidth{0pt}
\tablecolumns{8}
\tablecaption{Normalization and slope of the SZE-Temperature relations at $z$=0}
\tablehead{
\\
\multicolumn{1}{c}{$\Delta$}&
\multicolumn{1}{c}{} &
\multicolumn{1}{c}{$A_{5}^{\rm ad}$} &
\multicolumn{1}{c}{$\alpha_T^{\rm ad}$} &
\multicolumn{1}{c}{} &
\multicolumn{1}{c}{$A_{5}^{\rm csf}$} &
\multicolumn{1}{c}{$\alpha_T^{\rm csf}$} 
}
\startdata
\\
180m  & & $18.24^{+4.65}_{-3.71}$ & 2.52$\pm$0.18 & & $12.06^{+3.43}_{-2.67}$ & 2.60$\pm$0.21  \\
vir   & & $14.11^{+2.80}_{-2.34}$ & 2.52$\pm$0.15 & & $9.18^{+2.04}_{-1.67}$  & 2.63$\pm$0.19  \\
200c  & & $9.98^{+1.55}_{-1.34}$  & 2.50$\pm$0.13 & & $6.37^{+1.17}_{-0.99}$  & 2.62$\pm$0.17  \\
500c  & & $5.97^{+0.84}_{-0.74}$  & 2.45$\pm$0.12 & & $3.63^{+0.62}_{-0.53}$  & 2.61$\pm$0.17  \\
2500c & & $2.10^{+0.37}_{-0.32}$  & 2.50$\pm$0.17 & & $1.19^{+0.18}_{-0.16}$  & 2.73$\pm$0.19  \\
\enddata
%\normalsize
\label{tab:SZtm}
\end{deluxetable}
\vspace{0.8cm}
%%%%%%%%%%%%%%%%%%%%%%%%%%%%%%%%%%%%%%%%%%%%%%%%%%%%%%%%%

%---------------------------------------
\subsection{Comparisons to Previous Work}
\label{sec:comp}
%---------------------------------------

We now compare the results of our simulations to previous studies in
literature \citep{white_etal02,daSilva_etal04,motl_etal05}.  So far,
all of the previous studies have focused on simulating a large number
of clusters.  While these simulations provide good statistics, the
resolution is inevitably limited to capture relevant physical
processes. The present work is on the other end of the spectrum: the
resolution is high, but the statistic is low.  In this sense, our work
is complimentary to the previous studies.

Using SPH simulations, \citet{daSilva_etal04} studied the SZE scaling
relations in adiabatic simulations and re-simulations in which gas was
allowed to cool or was pre-heated.  These studies showed that both
strong gas cooling and preheating modify the slope of the SZE-M
relation.  They find that the slope in their cooling run
($\alpha_M=1.79$) was steeper than the slope of their adiabatic run
($\alpha_M=1.69$) by about 0.1.  More recently, \citet{motl_etal05}
used a large sample of clusters simulated using an Eulerian AMR code
and studied the SZE-M relation in the adiabatic run and three
resimulations in which they added gas cooling, star formation and
feedback one at a time.  These studies showed that the slope in the
simulations that include both gas cooling and star formation is very
similar to the slope of the adiabatic run.  Similar results are
obtained using simulations performed with the entropy conserving
Gadget code with cooling, star formation and feedback
\citep{white_etal02}.  Results of our simulations are also consistent
with these findings.  Moreover, \citet{motl_etal05} showed that the
slope in the cooling only run is steeper than the slope in the
adiabatic run by about 0.1, consistent with results reported by
\citet{daSilva_etal04}.

In addition, \citet{daSilva_etal04} investigated the redshift
evolution of the slope and scatter.  They showed that the slope and
normalization of the SZE-M relation evolve with redshift according to
the self-similar model out to $z$=2 in their adiabatic and cooling
only runs. \citet{motl_etal05} also found similar results in all four
sets of their simulations up to $z$=1.5.  We reached the same
conclusions using our high-resolution simulations.

In summary, the scatter, slope and redshift evolution of the SZE
observable-mass relation are generally insensitive to details of
cluster gas physics.  These results appear to be very robust.  In
fact, the agreement among different studies is rather remarkable,
because these simulations were carried out using very different
numerical techniques, resolution and implementation of various
physical processes incorporated in simulations.

Despite the robustness of the results discussed so far, the impact of
galaxy formation on normalization of the SZE scaling relations is not
yet well understood, because the effect is closely related to the cold
baryon fractions, which have not yet converged among different
simulations.  \citet{daSilva_etal04}, for example, find that the
normalization of the SZE-M relation in their cooling only run is lower
by 20\% and 9\% at $M_{200c}=10^{14}$ and $5\times
10^{14}h^{-1}M_{\odot}$ compared to the adiabatic run, respectively.
Moreover, \citet{white_etal02} find even smaller effects in their
simulation, because of the efficient energy feedback model used to
regulate conversion of hot gas into stars
\citep{springel_hernquist02}.  Using our high-resolution simulations,
we report that processes of galaxy formation lower the normalization
of the SZE-M relation by 41, 34 and 32\% at $r_{2500c}$, $r_{500c}$
and $r_{200c}$, respectively, almost uniformly throughout the cluster
mass range.  This is one of the largest impact of galaxy formation on
the SZE scaling relations reported to date.

% Fig.8: SZE-Mgas scaling relation
\begin{figure}[t]  
  \vspace{-0.0cm}
  \hspace{-0.0cm}
  \centerline{
    \psfig{figure=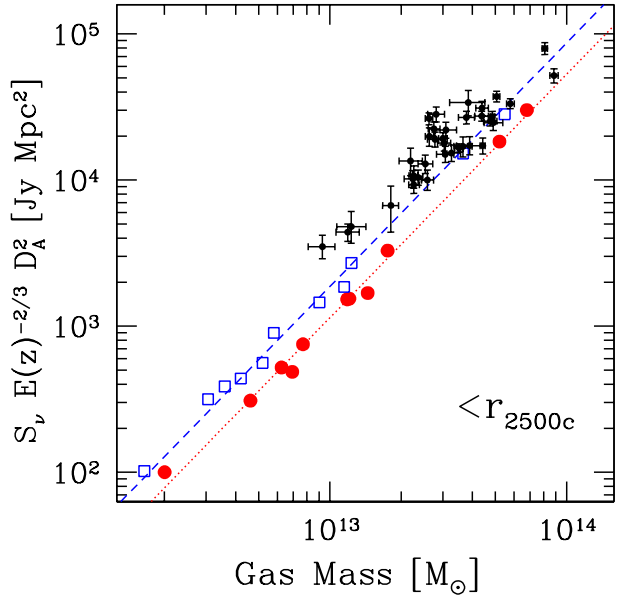,height=8.5cm}
  }
  \vspace{-0.2cm}
\caption{Comparisons of the SZ
flux-gas mass relations in the simulations and observations. The {\it
data points with errorbars} show the BIMA/OVRO SZ+Chandra X--ray
cluster observations for a sample of 36 clusters
\citep{laroque05}.  The SZ flux is the integrated flux within the
2D projected aperture of $R_{2500c}$, while the gas mass is the
enclosed gas mass within the sphere of $r_{2500c}$.  The SZ flux is
also corrected for the redshift evolution assuming the self-similar
evolution model and the $\Lambda$CDM cosmology.  The {\it solid
circles} and {\it open triangles} show the eleven simulated clusters
in the adiabatic and cooling and star formation runs, respectively, at
z=0.  The {\it dotted} and {\it dashed} lines are the best-fit
relations at z=0 of each set of simulation. The figure shows that the
simulations that include gas cooling and star formation are in
reasonably good agreement with the data, while the adiabatic
simulations are inconsistent with the observed relation.  }
\label{fig:szobs}
\end{figure}

%---------------------------------------
\subsection{Comparisons to Observations}
\label{sec:obs}
%---------------------------------------

To gauge how well the current simulations compare to data, we compare
the scaling relations in our simulations to recent observational
results. Figure~\ref{fig:szobs} shows a comparison of the SZE flux-gas
mass relation from our simulations to recent observational results
based on a sample of 36 clusters between $z$=0.14 and $z$=0.89,
observed with the SZE BIMA/OVRO interferometers and {\it Chandra}
X-ray telescope \citep{laroque05}. For comparison, we use the SZE
flux-gas mass relation, rather than the SZE-total mass relation,
because gas mass can be measured more accurately using {\it Chandra}
X-ray observations \citep[see][]{laroque05}.  The SZE flux is an
integrated flux within a 2D projected aperture of $R_{2500c}$, while
the gas mass is the enclosed gas mass within a sphere of $r_{2500c}$.
These quantities are computed using the best-fit parameters of the
isothermal $\beta$ model, fit jointly to the SZE and the {\it Chandra}
data \citep[see][for details]{laroque05}.  The observed SZE flux at a
given redshift is also corrected for the redshift evolution assuming
the canonical self-similar model and the $\Lambda$CDM cosmology.

In the simulations, the SZE flux is computed by projecting a sphere of
$3\times r_{500c}$ centered around the minimum of a cluster potential.
Note that hot gas in the cluster outskirt makes non-negligible
contribution to the integrated SZE signal, if the projected aperture
is a small fraction of the cluster virial radius.  For a projected
radius of $R_{2500c}$, approximately 35\% of the SZE signal arises
from a region outside a sphere of $r_{2500c}$, on average.  The
projection effect varies by about 20\% among different clusters, but
we did not find systematic variation of the effect with cluster mass.
The contribution of the cluster outskirt becomes less significant as
we make the projected aperture larger.  The projection of the hot gas
associated with dense structures in the foreground or background of
the cluster is not accounted for in this comparison.

The comparison shows that the adiabatic simulations are strongly
inconsistent with data.  Inclusion of gas cooling and star formation
brings simulations in a better agreement with observations. However, a
discrepancy between simulations and observations still remains.  More
specifically, there are indications that the normalization is larger,
and the slope is shallower for the observed relation
\citep[see][]{laroque05}.  However, the remaining discrepancy is at a
level of systematic uncertainties in current measurements \citep[see
e.g.,][]{benson_etal04}.  It is therefore critical to resolve
systematic uncertainties among different instruments and to increase
the cluster sample especially at low-masses to provide a better
leverage on the slope of the relations.

%-----------------------------------
\section{Discussion \& Conclusions}
\label{sec:discussion}
%-----------------------------------

We have presented the analysis of the Sunyaev-Zel'dovich effect (SZE)
scaling relations using high-resolution simulations of galaxy clusters
formed in a concordance $\Lambda$CDM cosmology.  The simulations of
eleven individual clusters spanning a decade in mass
($M_{500c}=3.5\times 10^{13}$ to $9\times 10^{14}h^{-1}M_{\odot}$) are
performed with the shock-capturing Eulerian adaptive mesh refinement
N-body+gasdynamics ART code.  We study the effects of gas cooling and
star formation on the SZE scaling relations and their redshift
evolution between $z$=0 and $z$=2 by comparing two sets of simulations
performed with and without these processes included.  The main results
are summarized as follows.

\newcounter{bean}
\begin{list}
{\arabic{bean}.}{ \usecounter{bean} \setlength{\parsep}{+0.03in}
\setlength{\leftmargin}{+0.15in} \setlength{\rightmargin}{+0.15in}}

\item{} The SZE signal integrated to a sufficiently large fraction of
cluster volume correlates very strongly with the enclosed total
cluster mass, independent of details of gas physics and dynamical
state of clusters.  The rms scatter of the SZE-total mass relation is
about 10-15\%.

\item{} The slope of the relation in the adiabatic run is in a very
good agreement with the predicted slope of the self-similar model in
the entire redshift range.  We find that the impact of galaxy
formation on the slope is small ($\lesssim 0.1$).

\item{} The redshift evolution of the SZE-total mass relation is
consistent with the self-similar model between $z$=0 and $z$=2: (a)
the slope is constant with redshift and (b) the normalization evolves
with redshift according to the self-similar evolution model.

\item{} Gas cooling and star formation significantly modify the
normalization of the SZE flux-total mass relation.  Inclusion of these
physical processes causes a decrease in the normalization by 41, 34
and 27\% at $r_{2500c}$, $r_{500c}$ and $r_{180m}$, respectively.  The
decrease is due to a large decrease in gas fraction, which is offset
somewhat by an increase in mass-weighted temperature.  Gas cooling and
star formation also cause an increase in total cluster mass and hence
modify the normalization by a few percent.

\item{} The integrated SZE signal also correlates strongly with gas
mass and mass-weighted temperature of clusters.  The results (1)-(3)
apply equally well for these relations, except that the redshift
evolution of the normalization exhibits some deviations from the
self-similar model, which increases toward higher redshifts.  Gas
cooling and star formation also significantly modify the normalization
of these relations.

\item{} The SZE flux-gas mass relation in the simulations with gas
cooling and star formation is in a better good agreement with the
observed relation for a sample of 36 OVRO/BIMA SZE+{\it Chandra} X-ray
observations \citep{laroque05} than the simulations neglecting galaxy
formation.

\end{list}

These results have a number of important implications for cosmological
studies with upcoming SZE cluster surveys.  First and foremost, the
SZE fluxes of clusters exhibit a remarkable regularity at all
redshifts, and the SZE signal integrated to a sufficiently large
fraction of cluster volume is insensitive to merging events \citep[see
e.g.,][]{motl_etal05} or properties of a cluster core (see
\S~\ref{sec:profiles}).  This indicates that the integrated SZE flux
is an extremely good proxy for cluster mass.  Second, the slope and
the redshift evolution of the SZE scaling relations are insensitive to
details of cluster gas physics, and they are well characterized by a
simple self-similar cluster model between $z$=0 and 2.  The simplicity
of their redshift evolution implies that the self-calibration
\citep{hu03,majumdar_etal04} will be effective.  Finally, these
results appear to be very robust, as the same conclusions have been
reached using simulations with very different numerical techniques,
resolution and implementation of various physical processes
incorporated in simulations (see \S~\ref{sec:comp}).

Despite the simplicity of redshift evolution, the normalization of the
SZE flux-mass relation is much less understood, because the effect is
closely related to the cold baryon fraction, which has not yet
converged among different simulations.  Using high-resolution cluster
simulations, we show that processes of galaxy formation have a
significant impact on the normalization of the SZE scaling relations.
Gas cooling and star formation suppress the normalization by $\approx
30-40\%$, primarily due to a large reduction in cluster gas mass
fraction.  Interestingly, the SZE scaling relations in these
simulations are in a reasonably good agreement with recent
observations.  Moreover, the gas mass fractions in these simulations
compare well with measurements from deep {\it Chandra} X--ray
observations of nearby clusters (Kravtsov et al. in
preparation). Despite these successes in matching SZE and X--ray
observations, the current cluster simulations may still suffer from
the "overcooling" problem since the fraction of baryons in the cold
gas and stars within a virial radius at $z$=0 is in a range
~0.25-0.35, at least a factor of two higher than observational
measurements \citep[see][and discussions and references
therein]{kravtsov_etal05}.

Given the importance of these issues, further efforts from observers
and theorists are needed to better understand the SZE scaling
relations for cosmological studies.  Observationally, it is important
to increase the sample size and the number of low-mass clusters.
Comparisons of different instruments will also help resolve systematic
uncertainties among different instruments \citep[see
e.g.,][]{benson_etal04}.  It is also critical to understand systematic
uncertainties in measurements of cluster mass through detailed and
extensive comparisons of X--ray, SZE and optical observations.  With
the advent of a number of dedicated SZE survey instruments, the
observational situation is expected to improve rapidly.

Theoretically, it is important to push detailed theoretical modelling
of cluster formation to investigate roles of various physical
processes, such as thermal conduction and AGN feedback, in shaping the
properties of the ICM.  For cosmological application, it is also
critical to understand projection effects \citep{white_etal02} and
nature of scatter in observable-mass relations \citep{lima_etal05}.
Numerical simulations of cluster formation will likely provide
important insights into these issues and help assess the effectiveness
of the self-calibration technique \citep{hu03,majumdar_etal04} for
future cluster surveys.

\acknowledgements

I would like to thank Andrey Kravtsov and John Carlstrom for their
invaluable guidance and advice during the course of this
Ph.D. project.  I would also like to thank Sam LaRoque for many useful
discussions on SZE observations and providing data points for the
comparison with simulations.  I also thank Gus Evrard, Wayne Hu and
Clem Pryke for useful feedback on this work and the anonymous referee
for many constructive comments on the manuscript.  This project was
supported by the NASA Graduate Student Researchers Program and by NASA
LTSA grant NAG5--7986. I also acknowledge the support from the Sherman
Fairchild Postdoctoral Fellowship at Caltech, where the final revision
to the manuscript was made.  The cosmological simulations used in this
study were performed on the IBM RS/6000 SP4 system ({\tt copper}) at
the National Center for Supercomputing Applications (NCSA).

%% FIGURES 

\end{document}